\documentclass{acmart}

\usepackage{braket}

\AtBeginDocument{%
  \providecommand\BibTeX{{%
    \normalfont B\kern-0.5em{\scshape i\kern-0.25em b}\kern-0.8em\TeX}}}

\copyrightyear{2023}
\acmYear{2023}
\setcopyright{acmlicensed}\acmConference[NANOARCH '23]{18th ACM International Symposium on Nanoscale Architectures}{December 18--20, 2023}{Dresden, Germany}
\acmBooktitle{18th ACM International Symposium on Nanoscale Architectures (NANOARCH '23), December 18--20, 2023, Dresden, Germany}
\acmPrice{15.00}
\acmDOI{10.1145/3611315.3633270}
\acmISBN{979-8-4007-0325-6/23/12}

\begin{document}

\title{A T-depth two Toffoli gate for 2D square lattice architectures}

\author{Alexandru Paler}
\affiliation{%
  \institution{Aalto University, Finland}
  \country{Finland}
}
\email{alexandru.paler@aalto.fi}

\author{Evan E. Dobbs, Joseph S. Friedman}
\affiliation{%
  \institution{The University of Texas at Dallas}
  \country{USA}
}

\begin{abstract}
We present a novel Clifford+T decomposition of a Toffoli gate. Our decomposition requires no SWAP gates in order to be implemented on 2D square lattices of qubits. This decomposition enables shallower, more fault-tolerant quantum computations on both NISQ and error-corrected architectures. We present the derivation of the circuit, and illustrate the qubit mapping on a Sycamore-like architecture.
\end{abstract}



\setlength{\belowcaptionskip}{-10pt}

\maketitle

\section{Introduction}

Practical quantum circuits, such as arithmetic circuits~\cite{paler2022realistic} or QRAM~\cite{paler2020parallelizing}, consist of Toffoli gates. However, most quantum computers cannot execute these gates natively, and the Toffoli gate is decomposed into sequences of hardware native gates. Very often the three-qubit Toffoli gate is expressed through a sequence of Clifford+T gates. The latter have increased compatibility with NISQ machines, and are also compatible with many quantum error-correcting codes -- for example, surface codes. Although the Toffoli gate is not considered NISQy, it is often used for developing and benchmarking co-design approaches (\textit{e.g.}, ~\cite{bowman2022hardware, niemann2020design}).

The feasibility of running quantum computations is a function of the circuit's depth and width. Architectural constraints, such as the available qubit connectivity, increase very often even by an order of magnitude the depth of the circuit~\cite{hu2019efficient, bowman2022hardware}. The lack of qubit all-to-all connectivity requires, for example, the insertion of SWAP gates for moving the qubit states~\cite{zulehner2018efficient}. However, the SWAP gates are not native, and are usually implemented by a sequence of three CNOTs. Therefore, it is not uncommon to notice even an order of magnitude cost (gate count, circuit depth) increase after adapting a quantum circuit to a given architecture. For this reason, architecture-aware compilation (\textit{e.g.}, ~\cite{hu2019efficient}) is a necessity for designing quantum circuits.

\begin{figure}[!t]
    \centering
    \includegraphics[width=0.7\columnwidth]{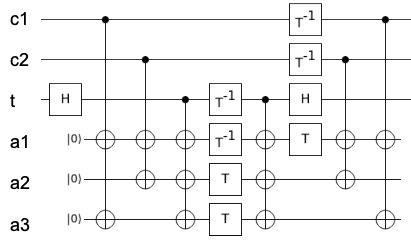}
    \caption{The T-depth two circuit that uses three ancillae is obtained by combining the Clifford+T circuits for an AND and controlled-S gate. The latter is applied on qubits c1 and a1 -- this is equivalent to applying it to the c1 and c2 qubits, due to how the ancillae are entangled to the control qubits. The first two CNOTs of Fig.~\ref{fig:cs}a do not need to be applied because the corresponding ancilla wire has already computed the necessary bit parity for the controlled-S gate phase polynomial. Similar T-depth two circuits, but with a single ancilla qubit, were presented in~\cite{amy2013meet, selinger2013quantum}.}
    \label{fig:combined}
\end{figure}

The cost of implementing a circuit can be expressed by the gate counts and circuit depth, as well as the T-depth, T-count, SWAP-depth, and SWAP-count. The T-count and the T-depth of a circuit are of interest when analysing error-corrected circuits, but not so much for NISQ circuits. One option for lowering the costs of designing architecture-aware quantum circuits is to use SWAP-free gate decompositions. This is the approach followed within this work.

In general, the Toffoli gate is decomposed into CNOTs and seven T gates. The decomposition gates can be arranged in a vast number of circuit configurations as long as a set of conditions is fulfilled; these conditions are expressed as phase polynomials \cite{selinger2013quantum}. One of the first Toffoli gate decompositions was presented in~\cite{barenco1995elementary}. The \emph{relative phase Toffoli} gate~\cite{maslov2016advantages}, which was initially described by Margolus, implements only up to a relative phase the Toffoli gate. This gate is also called the AND gate, and requires only four T gates in its decomposition; its phase polynomial has four terms. However, due to the relative phase, the AND gate cannot be easily used as a replacement for the Toffoli gate. Moreover, there exist complex trade offs when replacing Toffoli gates with ANDs and measurement-based uncomputations~\cite{paler2022realistic}. A Toffoli decomposition of 8 single-target (cannot be parallelized) CNOTs and depth 12 using no SWAP gates has been presented in~\cite{duckering2021orchestrated}.

We propose the novel Toffoli gate decomposition of 6 multi-target (paralellized) CNOTs and depth 8 which does not require SWAP gates when being executed on a 2D square lattice architecture (e.g. Google Sycamore quantum chip). We expect that our novel Toffoli gate will significantly lower the cost of implementing practical quantum computations. 

\section{Methods}
\begin{figure}[!h]
    \centering
    \includegraphics[width=0.6\columnwidth]{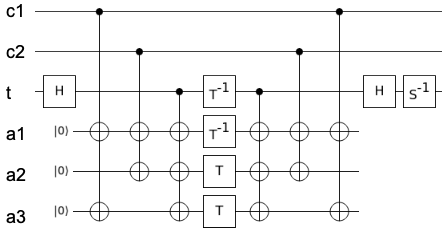}
    \caption{Logical AND gate with three ancillae and T-depth 1. The total depth is seven, because CNOTs share the same control wire and can be parallelised. The H and S gate can be executed in parallel with their neighbouring CNOTs.}
    \label{fig:and}
\end{figure}

\begin{figure}[!t]
    \centering
    \includegraphics[width=\columnwidth]{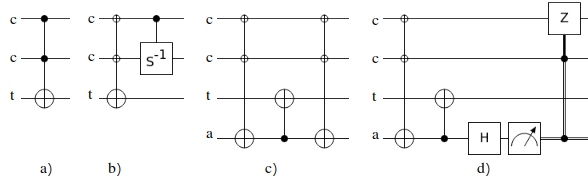}
    \caption{The conventional Toffoli (a) is implemented using: b) an AND gate and a controlled-S gate, c) an ancilla initialised in $\ket{0}$, two AND gates and a CNOT, d) an AND, an ancilla, a CNOT, and measurement-based uncomputation. The wire labels denote [c]ontrol, [t]arget, and [a]ncilla.}
    \label{fig:toff_and}
\end{figure}

We present a novel Toffoli gate (Fig.~\ref{fig:combined}) obtained by combining two methods: 1) building an AND gate similarly to the T-depth one Toffoli gate from ~\cite{jones2013low, selinger2013quantum} after placing four T gates according to the AND gate phase polynomial (Fig~\ref{fig:and}); 2) following the recipe from Fig. ~\ref{fig:toff_and} and adding a controlled-S gate next to the AND in order to build a Toffoli. Fig.~\ref{fig:cs} is the controlled-S gate of T-depth 1 using one ancilla. Fig. ~\ref{fig:combined} is the result of concatenating the Figs.~\ref{fig:and} and ~\ref{fig:cs}a. We have chosen T-depth two because a T-depth one would increase the degree of the required connectivity and the resulting circuit would not be compatible with a 2D square lattice.

\section{Preliminary Results}

\begin{figure} [!t]
    \centering
    \includegraphics[width=0.7\columnwidth]{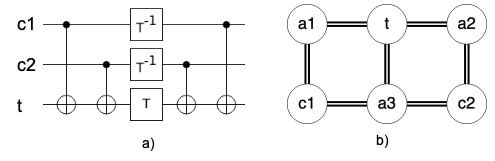}
    \caption{a) Clifford+T decomposition of control-S gate with one ancilla and T-depth; b) The circuit from Fig.~\ref{fig:combined} can be executed on a 2D square architecture without SWAPS. The vertices are qubits, the edges are the connections for implementing the CNOTs. The ancilla qubits $a_i$ are always targets of CNOTs, and the $c_1$, $c_2$ and $t$ are CNOT controls.}
    \label{fig:cs}
\end{figure}

The Toffoli gate from Fig.~\ref{fig:combined} requires no SWAP gates in order to be executed on a 2D square grid of qubits. Fig.~\ref{fig:cs}b illustrates the mapping of the Toffoli circuit qubits to the 2D grid. Each qubit from Fig.~\ref{fig:combined} is interacting with at most three other qubits, such that it is possible to map the Clifford+T decomposition to a square grid. If any of the qubits would have been interacted with four other qubits, the layout could have still been 2D, but it would have been absolutely necessary to use SWAP gates to implement multi-Toffoli circuits: for example, the $t$ qubit from Fig.~\ref{fig:cs}b could be used for another Toffoli gate after tiling the layout in 2D. If the same $t$ qubit would have been interacted with four other qubits, then tiling would not be possible without using SWAP gates.

The presented Toffoli gate can be used to reduce the cost of designing achitecture-aware computations such as the ones presented in~\cite{hu2019efficient}. Future work will focus on implementing full algorithms and quantifying the cost reduction after using our new gate. Moreover, we will analyze SWAP-free Toffoli decompositions for the heavy hex layout of the IBM quantum chips.

\begin{acks}
This research was developed in part with funding from the Defense Advanced Research Projects Agency [under the Quantum Benchmarking (QB) program under award no. HR00112230007 and HR001121S0026 contracts]. 
\end{acks}

\bibliographystyle{ACM-Reference-Format}
\bibliography{__main}


\begin{thebibliography}{12}


\ifx \showCODEN    \undefined \def \showCODEN     #1{\unskip}     \fi
\ifx \showDOI      \undefined \def \showDOI       #1{#1}\fi
\ifx \showISBNx    \undefined \def \showISBNx     #1{\unskip}     \fi
\ifx \showISBNxiii \undefined \def \showISBNxiii  #1{\unskip}     \fi
\ifx \showISSN     \undefined \def \showISSN      #1{\unskip}     \fi
\ifx \showLCCN     \undefined \def \showLCCN      #1{\unskip}     \fi
\ifx \shownote     \undefined \def \shownote      #1{#1}          \fi
\ifx \showarticletitle \undefined \def \showarticletitle #1{#1}   \fi
\ifx \showURL      \undefined \def \showURL       {\relax}        \fi
\providecommand\bibfield[2]{#2}
\providecommand\bibinfo[2]{#2}
\providecommand\natexlab[1]{#1}
\providecommand\showeprint[2][]{arXiv:#2}

\bibitem[Amy et~al\mbox{.}(2013)]%
        {amy2013meet}
\bibfield{author}{\bibinfo{person}{Matthew Amy}, \bibinfo{person}{Dmitri Maslov}, \bibinfo{person}{Michele Mosca}, {and} \bibinfo{person}{Martin Roetteler}.} \bibinfo{year}{2013}\natexlab{}.
\newblock \showarticletitle{A meet-in-the-middle algorithm for fast synthesis of depth-optimal quantum circuits}.
\newblock \bibinfo{journal}{\emph{IEEE Transactions on Computer-Aided Design of Integrated Circuits and Systems}} \bibinfo{volume}{32}, \bibinfo{number}{6} (\bibinfo{year}{2013}), \bibinfo{pages}{818--830}.
\newblock


\bibitem[Barenco et~al\mbox{.}(1995)]%
        {barenco1995elementary}
\bibfield{author}{\bibinfo{person}{Adriano Barenco}, \bibinfo{person}{Charles~H Bennett}, \bibinfo{person}{Richard Cleve}, \bibinfo{person}{David~P DiVincenzo}, \bibinfo{person}{Norman Margolus}, \bibinfo{person}{Peter Shor}, \bibinfo{person}{Tycho Sleator}, \bibinfo{person}{John~A Smolin}, {and} \bibinfo{person}{Harald Weinfurter}.} \bibinfo{year}{1995}\natexlab{}.
\newblock \showarticletitle{Elementary gates for quantum computation}.
\newblock \bibinfo{journal}{\emph{Physical review A}} \bibinfo{volume}{52}, \bibinfo{number}{5} (\bibinfo{year}{1995}), \bibinfo{pages}{3457}.
\newblock


\bibitem[Bowman et~al\mbox{.}(2022)]%
        {bowman2022hardware}
\bibfield{author}{\bibinfo{person}{Max~Aksel Bowman}, \bibinfo{person}{Pranav Gokhale}, \bibinfo{person}{Jeffrey Larson}, \bibinfo{person}{Ji Liu}, {and} \bibinfo{person}{Martin Suchara}.} \bibinfo{year}{2022}\natexlab{}.
\newblock \showarticletitle{Hardware-Conscious Optimization of the Quantum Toffoli Gate}.
\newblock \bibinfo{journal}{\emph{arXiv preprint arXiv:2209.02669}} (\bibinfo{year}{2022}).
\newblock


\bibitem[Duckering et~al\mbox{.}(2021)]%
        {duckering2021orchestrated}
\bibfield{author}{\bibinfo{person}{Casey Duckering}, \bibinfo{person}{Jonathan~M Baker}, \bibinfo{person}{Andrew Litteken}, {and} \bibinfo{person}{Frederic~T Chong}.} \bibinfo{year}{2021}\natexlab{}.
\newblock \showarticletitle{Orchestrated trios: compiling for efficient communication in quantum programs with 3-qubit gates}. In \bibinfo{booktitle}{\emph{Proceedings of the 26th ACM International Conference on Architectural Support for Programming Languages and Operating Systems}}. \bibinfo{pages}{375--385}.
\newblock


\bibitem[Hu et~al\mbox{.}(2019)]%
        {hu2019efficient}
\bibfield{author}{\bibinfo{person}{Shaohan Hu}, \bibinfo{person}{Dmitri Maslov}, \bibinfo{person}{Marco Pistoia}, {and} \bibinfo{person}{Jay Gambetta}.} \bibinfo{year}{2019}\natexlab{}.
\newblock \showarticletitle{Efficient circuits for quantum search over 2D square lattice architecture}. In \bibinfo{booktitle}{\emph{Proceedings of the 56th Annual Design Automation Conference 2019}}. \bibinfo{pages}{1--2}.
\newblock


\bibitem[Jones(2013)]%
        {jones2013low}
\bibfield{author}{\bibinfo{person}{Cody Jones}.} \bibinfo{year}{2013}\natexlab{}.
\newblock \showarticletitle{Low-overhead constructions for the fault-tolerant Toffoli gate}.
\newblock \bibinfo{journal}{\emph{Physical Review A}} \bibinfo{volume}{87}, \bibinfo{number}{2} (\bibinfo{year}{2013}), \bibinfo{pages}{022328}.
\newblock


\bibitem[Maslov(2016)]%
        {maslov2016advantages}
\bibfield{author}{\bibinfo{person}{Dmitri Maslov}.} \bibinfo{year}{2016}\natexlab{}.
\newblock \showarticletitle{Advantages of using relative-phase Toffoli gates with an application to multiple control Toffoli optimization}.
\newblock \bibinfo{journal}{\emph{Physical Review A}} \bibinfo{volume}{93}, \bibinfo{number}{2} (\bibinfo{year}{2016}), \bibinfo{pages}{022311}.
\newblock


\bibitem[Niemann et~al\mbox{.}(2020)]%
        {niemann2020design}
\bibfield{author}{\bibinfo{person}{Philipp Niemann}, \bibinfo{person}{Alexandre~AA de Almeida}, \bibinfo{person}{Gerhard Dueck}, {and} \bibinfo{person}{Rolf Drechsler}.} \bibinfo{year}{2020}\natexlab{}.
\newblock \showarticletitle{Design space exploration in the mapping of reversible circuits to ibm quantum computers}. In \bibinfo{booktitle}{\emph{2020 23rd Euromicro Conference on Digital System Design (DSD)}}. IEEE, \bibinfo{pages}{401--407}.
\newblock


\bibitem[Paler et~al\mbox{.}(2020)]%
        {paler2020parallelizing}
\bibfield{author}{\bibinfo{person}{Alexandru Paler}, \bibinfo{person}{Oumarou Oumarou}, {and} \bibinfo{person}{Robert Basmadjian}.} \bibinfo{year}{2020}\natexlab{}.
\newblock \showarticletitle{Parallelizing the queries in a bucket-brigade quantum random access memory}.
\newblock \bibinfo{journal}{\emph{Physical Review A}} \bibinfo{volume}{102}, \bibinfo{number}{3} (\bibinfo{year}{2020}), \bibinfo{pages}{032608}.
\newblock


\bibitem[Paler et~al\mbox{.}(2022)]%
        {paler2022realistic}
\bibfield{author}{\bibinfo{person}{Alexandru Paler}, \bibinfo{person}{Oumarou Oumarou}, {and} \bibinfo{person}{Robert Basmadjian}.} \bibinfo{year}{2022}\natexlab{}.
\newblock \showarticletitle{On the Realistic Worst-Case Analysis of Quantum Arithmetic Circuits}.
\newblock \bibinfo{journal}{\emph{IEEE Transactions on Quantum Engineering}}  \bibinfo{volume}{3} (\bibinfo{year}{2022}), \bibinfo{pages}{1--11}.
\newblock


\bibitem[Selinger(2013)]%
        {selinger2013quantum}
\bibfield{author}{\bibinfo{person}{Peter Selinger}.} \bibinfo{year}{2013}\natexlab{}.
\newblock \showarticletitle{Quantum circuits of T-depth one}.
\newblock \bibinfo{journal}{\emph{Physical Review A}} \bibinfo{volume}{87}, \bibinfo{number}{4} (\bibinfo{year}{2013}), \bibinfo{pages}{042302}.
\newblock


\bibitem[Zulehner et~al\mbox{.}(2018)]%
        {zulehner2018efficient}
\bibfield{author}{\bibinfo{person}{Alwin Zulehner}, \bibinfo{person}{Alexandru Paler}, {and} \bibinfo{person}{Robert Wille}.} \bibinfo{year}{2018}\natexlab{}.
\newblock \showarticletitle{An efficient methodology for mapping quantum circuits to the IBM QX architectures}.
\newblock \bibinfo{journal}{\emph{IEEE Transactions on Computer-Aided Design of Integrated Circuits and Systems}} \bibinfo{volume}{38}, \bibinfo{number}{7} (\bibinfo{year}{2018}), \bibinfo{pages}{1226--1236}.
\newblock


\end{thebibliography}

\end{document}